\documentclass[twocolumn,showpacs]{revtex4}
\usepackage{epsfig}
\usepackage[english]{babel}
\usepackage{latexsym}
\usepackage{graphics}
\usepackage{subfigure}
\usepackage{epsfig}
\usepackage{graphicx}
\usepackage{dcolumn}
\usepackage{amsmath}

\begin{document}

\title{{Ionization Suppression of Diatomic Molecules in Intense Midinfrared Laser Field}
\footnotetext{$^{*}$ xjliu@wipm.ac.cn}

\footnotetext{$^{\dag}$ chen$\_$jing@iapcm.ac.cn}

\footnotetext{$^{\ddag}$ ycheng-45277@hotmail.com}

\footnotetext{$^{\S}$ zzxu@mail.shcnc.ac.cn}
 }

\author{Z. Lin$^{1,6}$, X. Y. Jia$^{2,3}$, C. Wang$^{1,6}$, Z. Hu$^{1,6}$, H. Kang$^{1}$, W. Quan$^{1}$, X.Y. Lai$^{1}$, X. Liu$^{1,*}$, J. Chen$^{4,5,\dag}$,
B. Zeng$^{2}$, W. Chu$^{2}$, J. P. Yao$^{2}$, Y. Cheng$^{2,\ddag}$,
Z. Z. Xu$^{2,\S}$}

\begin{abstract}
Diatomic molecules (e.g., O$_2$) in intense laser field exhibit a
peculiar suppressed ionization behavior compared to their companion
atoms. Several physical models have been proposed to account for
this suppression while no consensus has been achieved. In this
letter, we aim to clarify the underlying mechanisms behind this
molecular ionization suppression. Experimental data recorded at
midinfrared laser wavelength and its comparison with that at
near-infrared wavelength revealed a peculiar wavelength and
intensity dependence of the suppressed ionization of O$_2$ with
respect to its companion atom of Xe, while N$_2$ behaves like a
structureless atom. It is found that the S-matrix theory calculation
can reproduce well the experimental observations and unambiguously
identifies the significant role of two-center interference effect in
the ionization suppression of O$_2$.
\end{abstract}

\affiliation{$^{1}$State Key Laboratory of Magnetic Resonance and
Atomic and Molecular Physics, Wuhan Institute of Physics and
Mathematics, Chinese Academy of Sciences, Wuhan 430071,
China\\$^{2}$State Key Laboratory of High Field Laser Physics,
Shanghai Institute of Optics and Fine Mechanics, Chinese Academy
of Sciences, P.O.Box 800-211, Shanghai 201800,
China\\$^{3}$Quantum Optoelectronics Laboratory, Southwest
Jiaotong University, Chengdu 610031, China\\$^{4}$Center for
Applied Physics and Technology, Peking University, Beijing 100084,
China\\$^{5}$Institute of Applied Physics and Computational
Mathematics, P. O. Box 8009, Beijing 100088, China\\$^{6}$Graduate
School of Chinese Academy of Sciences, Beijing 100080, China\\}

 \maketitle

Ionization of molecules in intense laser pulses plays a central role
in understanding strong field molecular physics. Most molecular
strong field processes, such as above-threshold ionization (ATI),
high-harmonic generation (HHG), double ionization (DI), and Coulomb
explosion (CE) are derived directly from this fundamental process.
Compared to atoms, for which the ionization and related phenomena
have been well understood \cite{Dimauro1995AAMOP}, molecular
ionization exhibits a large variety of peculiar behaviors due to its
structural complexity and the extra nuclear degrees of freedom
\cite{Posthumus2004RPP}.

One of the most fundamental differences between molecule and its
companion atom, i.e., the atom with a comparable ionization
potential, is represented by their relatively distinct ionization
probability. For example, when subject to a Ti:Sapphire laser pulse
at $\sim$ 800 nm, a strong suppression has been observed in
ionization probability of diatomic molecule O$_2$ compared to the
rare gas atom Xe, while no suppression is seen in the diatomic
molecule N$_2$ compared to its companion atom Ar
\cite{Chin1996JPB,Guo1998PRA}. Several theoretical models, including
the KFR (or Keldysh-Faisal-Reiss)
\cite{Becker2000PRL,Becker2004PRA}, multielectron screening
\cite{Guo2000PRL} or MO-ADK \cite{Tong2002PRA} models, have been
proposed to address this issue. The KFR model predicts that the
interference between ionizing wave packets emitted from the two
distinct nuclear centers can lead to ionization suppression for
molecules (e.g., O$_2$) with antisymmetric electronic ground states.
The multielectron screening model introduces a charge-screening
correction to the tunneling theory. In the MO-ADK model, the
difference between atomic and molecular ionization is attributed to
different asymptotic behaviors of their ground state wave functions
\cite{Tong2002PRA}. Though all the proposed models can, in
principle, account for the experimental observations, mostly at
near-infrared wavelength of 800 nm, no consensus on the underlying
mechanism has been achieved so far. In addition, various \emph{ab
initio} methods have been applied to study this problem and
qualitative agreement with the experimental results has been
achieved \cite{CC2004PRA,DR2005PRA,TC2009PRA}, however, no clear
physical mechanism can be identified. Recently, alignment dependence
of molecular ionization has been measured and the comparison with
theories \cite{PLRCV2007PRL} shed more light on the underlying
molecule specific effects.

Further understanding and clarification of the mechanism behind the
distinct molecular ionization requires, from experimental point of
view, the extension of the measurements into other wavelength range
than the solely 800 nm. Indeed, recent experiment performed at a
shorter wavelength of 400 nm \cite{Wu2006PRL} exhibited a similar
behavior of O$_2$, and, however, a rather disparate behavior of
N$_2$. The latter is found to have a higher ionization probability
compared to that of Ar for linear polarization while the difference
vanishes for circular polarization. This has been explained by the
resonance enhancement, a characteristic of multiphoton ionization
(MPI) process, in N$_2$. On the other side, this MPI resonance
effect may also contribute significantly to the atomic or molecular
ionization at 800 nm \cite{atiresonance}, frustrating an explicit
comparison with the theory and a clear identification of the
mechanism.

In this letter, we perform a comparison study of intense field
ionization between diatomic molecules (i.e., N$_2$ and O$_2$) and
their companion atoms (i.e., Ar and Xe) at a midinfrared wavelength
of 2000 nm. In contrast to previous studies carried out at shorter
laser wavelengths, our experiment at long wavelength ensures that
the ionization process falls deeply within the tunnel ionization
(TI) regime \cite{Blaga2009NatPhys,Quan2009PRL}. The comparison of
the data with the theory provides a crucial clue to the physical
mechanism behind molecular ionization in TI regime, which is of
special importance to the emerging field of ultrafast imaging of
molecular structure and dynamics, wherein the ionized electrons from
the molecules can be employed as a tool in imaging of molecular
orbitals \cite{Itatani2004Nature,Meckel2008Science,Kang2010PRL} and
probing nuclear dynamics with attosecond resolution
\cite{Baker2006Science}.

In our experiments, wavelength-tunable mid-infrared femtosecond
laser pulses are generated by an optical parametric amplifier (OPA,
TOPAS-C, Light Conversion, Inc.) pumped by a commercial Ti:Sapphire
laser system (Legend, Coherent, Inc.). This OPA system has been
described in detail elsewhere \cite{fu}. The pulse energy from OPA
is variable, before focused into the vacuum chamber, by means of an
achromatic half-wave plate followed by a polarizer. A standard
time-of-flight (TOF) mass spectrometer is used to register the ion
signal. By means of a cryopump, the base pressure of the
spectrometer is maintained below $10^{-8}$ mbar. At the end of the
spectrometer, ions are detected with a microchannel plate as a
function of flight time. The ion signal is further amplified,
discriminated, and sent to a multihit time digitizer to generate TOF
mass spectra. Depending on the laser intensity, the data point in
the ion yield plots is obtained by averaging over $10^4$ up to $6
\times 10^6$ laser shots at each intensity to ensure a sufficiently
high statistical accuracy.

\begin{figure}[t]
\begin{center}
\rotatebox{0}{\resizebox *{9.0cm}{7.5cm}
{\includegraphics{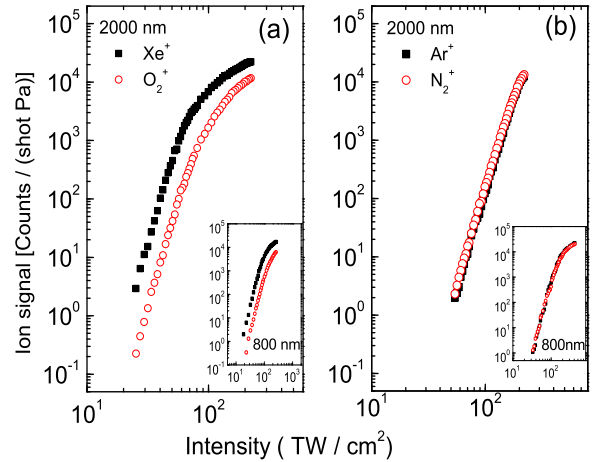}}}
\end{center}
\caption{(Color online). Experimental single ionization yields of
diatomic molecules (i.e., N$_2$ and O$_2$) and their companion atoms
(i.e., Ar and Xe) as a function of laser intensity at 2000 nm. The
corresponding data recorded at 800 nm are also shown in the inset
for comparison.} \label{fig1}
\end{figure}

In Figs. \ref{fig1} (a) and (b), we present the measured ion yields
of singly charged O$_2$ versus Xe and N$_2$ versus Ar, respectively,
using a linearly polarized light at a center wavelength of 2000 nm.
For comparison, data recorded at Ti:Sapphire laser wavelength of 800
nm are also shown in the inset. Very similar to previous studies at
only 800 nm \cite{Guo1998PRA}, our data show that N$_2$ and Ar have
a parallel ionization probabilities also at 2000 nm. While for
O$_2$, a significant suppression of ionization yield compared to Xe
is found for both 800 and 2000 nm fields.

\begin{figure}[t]
\begin{center}
\rotatebox{0}{\resizebox *{9.0cm}{6.0cm} {\includegraphics
{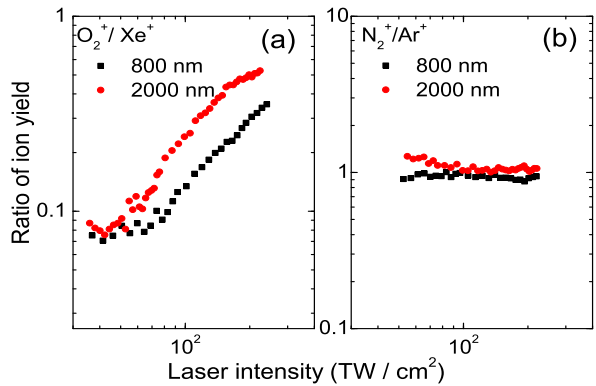}}}
\end{center}
\caption{ (Color online). Experimental ratios of ionization yield
between molecules and their companion atoms for both 800 and 2000
nm, as a function of laser intensity.} \label{fig3}
\end{figure}

In order to compare quantitatively the relative ion yields of the
molecules with respect to their companion atoms and to provide
more information for the benefit of understanding the underlying
mechanism, we plot the ratio of ionization yield of O$_2$ with
respect to Xe in Fig. \ref{fig3}(a), and N$_2$ with respect to Ar
in Fig. \ref{fig3}(b), respectively. It is found that the ratio of
N$_2^+$/Ar$^+$ keeps almost constant, to be around 1, suggesting
that the N$_2$ always behaves like a structureless atom,
irrespective of the laser intensity and wavelength. On the other
side, the ratio of O$_2^+$/Xe$^+$ shows a strong dependence on the
laser intensity. The ratio gradually increases with increasing
intensity in both 800 nm and 2000 nm cases. More interestingly,
the ratio shows a clear wavelength dependence. Two ratios almost
coincide at about the lowest intensity used in the experiment but
the ratio of 2000 nm apparently increases faster with intensity
than that of 800 nm, which indicates that, for the same
intensities above $5\sim6\times 10^{13}$W/cm$^2$, the longer the
wavelength, the less pronounced ionization suppression of O$_2$
compared to Xe.

This apparent dependence of the ratio of O$_2^+$/Xe$^+$ on the laser
wavelength is inconsistent with the MO-ADK model prediction
\cite{Tong2002PRA} since ADK-type tunneling formulation is based on
a quasi-static approximation which will give an ionization rate
independent of the wavelength. On the other side, the multielectron
screening model predicts that the suppression of the O$_2$ compared
to Xe should be more pronounced with the increase of the laser
wavelength \cite{Guo2000PRL,Wu2006PRL}, which is also in
disagreement with the experimental observations. In contrast, we
will see below that this dependence of the ratio of O$_2^+$/Xe$^+$
on both the wavelength and the intensity can be well understood
within the S-matrix formulation and is closely related to the
destructive interference of the two subwaves of the ionizing
electron emerging from the two atomic centers of O$_2$ with ground
state of antibonding symmetry.

In the S-matrix theory, single ionization rate for an atom or a
molecule in a linearly polarized laser field with a vector potential
$\mathbf{A}(t)=A_0 \mathbf{e}_z\cos(\omega t)$ is given by
\cite{Becker2000PRL,Keldysh}
\begin{align}
W=&2\pi N_e \sum\limits_{N=N_0} k_N(U_p-N\omega)^2 \int
d\widehat{k}_N J_N^2(\mathbf{\mathbf{k_N}\cdot\mathbf{\alpha}},
\frac{U_p}{2\omega})\nonumber\\
&\times|\langle\phi_{k_N}(\mathbf{r})|\psi_i(\mathbf{r})\rangle|^2.
\label{1a}
\end{align}
Here $N_e$ denotes the number of equivalent electrons and $N_0$ is
the minimum number of the photons needed to ionize the target.
$k_N=\sqrt{2(N\omega-I_p-U_p)}$ and $\textbf{k}_N $ represents the
momentum of the emitted electron, with $I_p$ and $U_p$ the
ionization potential and ponderomotive energy, respectively. $J_N$
is a generalized Bessel function with
$\mathbf{\alpha}=A_0\mathbf{e}_z/\omega$. $\phi_{k_N}(\mathbf{r})$
denotes the plane wave function and $\psi_i(\mathbf{r})$ is the
ground state wave function of the atom or molecule. In our
calculation, the wave function of atomic ground state is
approximated by the outmost single electron orbital, while for
molecules we use the linear combination of atomic orbitals to
simulate the molecular orbital (LCAO-MO) approximately
\cite{Usachenko2005PRA, Levine}. Moreover, according to \emph{ab
initio} calculation using time-dependent density-functional theory
in Ref. \cite{TC2009PRA}, the highest-occupied molecular orbital
(HOMO) (1$\pi_g$) dominates in the ionization process of O$_2$
molecule while both HOMO (3$\sigma_g$) and HOMO-1 (1$\pi_u$) play
important role for N$_2$. Therefore, only HOMO is considered for
O$_2$ and both HOMO and HOMO-1 are considered for N$_2$ in our
calculation.

\begin{figure}[t]
\begin{center}
\rotatebox{0}{\resizebox *{9.0cm}{6.0cm} {\includegraphics
{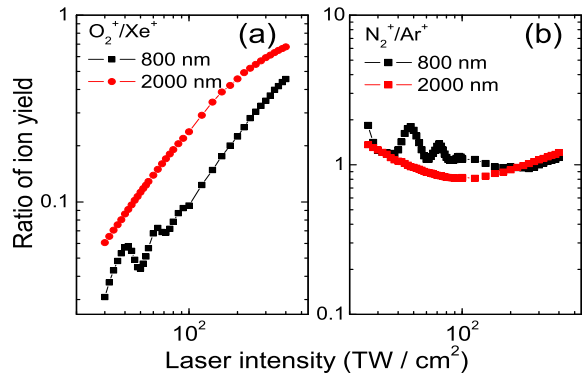}}}
\end{center}
\caption{ (Color online). Ratios of ionization yield between
molecules and their companion atoms for both 800 and 2000 nm,
calculated from the S-matrix theory, as a function of laser
intensity.} \label{fig4}
\end{figure}

The calculated ratio of single ionization rates for molecules
O$_2$ and N$_2$ with respect to their companion atoms Xe and Ar as
a function of laser intensity at 800 nm and 2000 nm are shown in
Figs. \ref{fig4}(a) and (b), respectively. It is found that the
theoretical results are well consistent with the experimental data
\cite{note1}. The ratios for N$_2^+$/Ar$^+$ keeps around 1,
irrespective of the field intensity and wavelength. In contrast,
the ratios for O$_2^+$/Xe$^+$ show strong suppression and raise
with the increase of the laser intensity. Moreover, the ratio of
2000 nm keeps higher than that of 800 nm for the whole intensity
regime considered. Especially, both experimental and theoretical
data agree on that the ratio of O$_2^+$/Xe$^+$ depends linearly on
the laser intensity (i.e., $\propto I$), e.g., the ratio of 800 nm
increases by about 4 times when the laser intensity increases by
about 4 times (see Fig. \ref{fig3}(a)). It is noteworthy that this
amount of increase is significantly larger than that predicted by
the MO-ADK theory which gives a scaling of $I^{1/2}$
\cite{Tong2002PRA}.

Further comprehension of the distinct intensity and wavelength
dependence of the ratio of O$_2^+$/Xe$^+$ can be drawn from the
S-matrix formula, in which the single ionization rate of molecules
consists of two major parts that comes from the Fourier transform of
the ground state. For O$_2$, the Fourier transform has the form
$\Psi_i(\textbf{k}_N)=C\Phi_{at}(\textbf{k}_N)\sin (\mathbf{k}_N
\cdot \mathbf{R}/2)$ with $C$ being the normalization factor. Here
$\Phi_{at}(\textbf{k}_N)$ is the atom-like part, which is determined
mainly by the atomic orbital constituting the molecular orbital
($2p_x$ wavefunction for O$_2$), and the trigonometric part $\sin
(\mathbf{k}_N \cdot \mathbf{R}/2)$ depends on the molecular
structure and is associated with the interference effect between the
wavepackets of the ionizing electrons centered at the individual
nucleus \cite{Becker2000PRL}. This trigonometric part leads to the
suppression effect since it always gives destructive interference
when $\mathbf{k}_N \cdot \mathbf{R}\ll\pi$ which is usually
satisfied in the conditions considered in this work.

\begin{figure}[t]
\begin{center}
\rotatebox{0}{\resizebox *{9.0cm}{8.0cm} {\includegraphics
{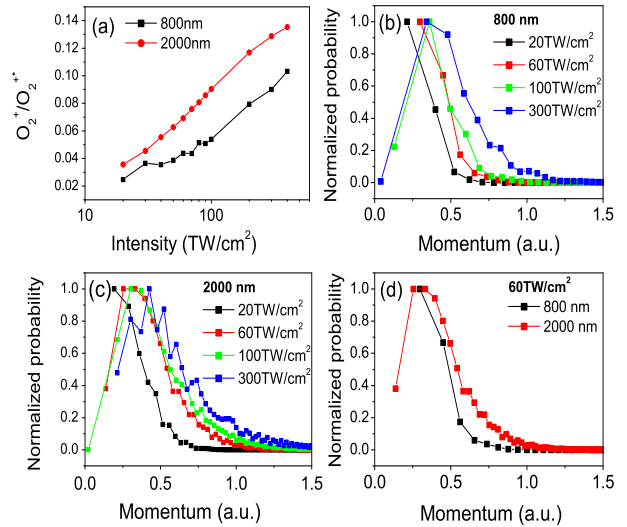}}}
\end{center}
\caption{(Color online). (a): Calculated ratio between ionization
yield of O$_2$ with (denoted as O$_2^{+}$) and without the
trigonometric term (denoted as O$_2^{+*}$) (see text). (b), (c) and
(d): Normalized momentum spectrum of the photoelectron from O$_2$
calculated without the trigonometric term at molecular alignment
$45^o$ with respect to the laser field. The field parameters used in
the calculations are shown in the panels.} \label{fig5}
\end{figure}

This suppression effect can be clearly seen in Fig. \ref{fig5}(a)
which shows the calculated ion yield ratio between the two cases
with and without the trigonometric term included in Eq. (\ref{1a}),
denoted by O$_2^+$ and O$_2^{+*}$, respectively, for both 800 nm and
2000 nm laser wavelengths. When the trigonometric term is included,
the ionization yield is significantly suppressed. Moreover, the
ratio ascends with intensity and increases with laser wavelength. To
understand these effects, we plot the momentum spectra of
photoelectrons emitted from the ionization of molecular O$_2$,
calculated without the trigonometric term, for different laser
intensities and wavelengths. As shown in Figs. \ref{fig5}(b) and
(c), calculated at 800 and 2000 nm, respectively, the contribution
from large $k_N$ becomes more significant as the intensity
increases. This will cause an increase of the interference term
$\sin (\mathbf{k}_N \cdot \mathbf{R}/2)$ and as a consequence, the
ascending ratio with the increase of the laser intensity, in
agreement with the results in Fig. \ref{fig5}(a). In addition, for a
certain laser intensity the momentum spectrum becomes broader when
the wavelength increases (see Fig. \ref{fig5}(d)). Similar to the
intensity effect, this results in an increasing ratio and less
pronounced suppression effect for longer wavelength, which is
consistent well with the experiment. Note that in Figs.
\ref{fig5}(b), (c) and (d), for simplicity, we only show the
electron distribution for a specific molecular alignment angle of
$45^o$ with respect to field direction, since the ionization rate of
O$_2$ reaches maximum at about this angle \cite{Kang2010PRL}. It is
worthy to mention that the distributions are not sensitive to the
alignment angle in our calculations. Therefore, the interference
effect from the trigonometric term originating from the multi-center
feature of molecules plays an essential role in the ionization
suppression of O$_2$ molecule. For N$_2$ molecule, the situation is
much more complicated. Besides the HOMO orbital, the HOMO-1 orbital
contributes to the total ion yield and is becoming especially
important at high intensities. Moreover, the HOMO orbital is an
admixture of both atomic $s$ and $p$ orbitals. The contributions of
these two types of orbitals, which possess trigonometric terms of
both $\cos (\mathbf{k}_N \cdot \mathbf{R}/2)$ and $\sin
(\mathbf{k}_N \cdot \mathbf{R}/2)$, add coherently in the
calculation \cite{Madsen2006PRA}. As a result, N$_2^+$/Ar$^+$ shows
no suppression and is hardly dependent on the laser wavelength.

Note that the ratio of O$_2^+$/Xe$^+$ has a scaling of $\propto I$
dependence on the laser intensity (see Figs. \ref{fig3}(a) and
\ref{fig4}(a)), in contrast to that the ratio shown in Fig.
\ref{fig5}(a), introduced solely by the interference effect, gives
a scaling of $\propto I^{1/2}$. The other contribution accounting
for the additional $I^{1/2}$ comes from the effect of the atomic
orbitals. It is noteworthy that the outmost orbital of O$_2$
molecule is composed of a $2p_x$ orbital while the outmost orbital
of xenon atom is $5p_z$. Our calculation shows that the ratio
between the ionization yield from these two orbitals will also
increase by about 3$\sim$4 times when the intensity increases by
about one order of magnitude, which gives an additional $I^{1/2}$
dependence of the ratio O$_2^+$/Xe$^+$. Therefore, the two-center
interference effect, together with the atomic orbital effect,
gives rise to the experimentally observed intensity dependence
($\propto I$) of the O$_2^+$/Xe$^+$.

A closer comparison between our theoretical simulation (Fig.
\ref{fig4}(a)) and experimental data (Fig. \ref{fig3}(a)) shows a
perceptible discrepancy at very low intensity regime. The
experimental ratio does not decrease with decreasing intensity while
the theoretical ratio keeps dropping at about $4\sim5\times 10^{13}$
W/cm$^2$. This discrepancy may be partially attributed to resonance
effect that becomes important when the laser intensity is low and
the ionization is well in the multiphoton regime \cite{Wu2006PRL}.
In the multiphoton regime, the ionization channel via
Freeman-resonance process \cite{Freeman1987PRL} contributes
significantly to the ionization yield. It is well-known that O$_2$
molecule possesses more abundant highly excited states than Xe and
hence provides more resonance channels for the ATI process,
resulting in the increased ratio of O$_2^+$/Xe$^+$ comparing with
that given by the S-matrix calculation in which all the resonance
processes are ignored.

In summary, we present a comparison study on the ionization of
diatomic molecules (N$_2$ and O$_2$) and their companion atoms (Ar
and Xe) at midinfrared wavelength. Our experimental data reveals
that the ionization probability of N$_2$ is almost identical to its
companion atom of Ar. In contrast, O$_2$ exhibits a distinct
suppression compared to Xe and more importantly, a strong dependence
of this suppression on both the laser wavelength and intensity has
been found. While this finding is in conflict with the molecular ADK
formulation and multielectron screening model predictions, it can be
well reproduced by the S-matrix theory calculation, which considers
the interference between ionizing wave packets emitted from the two
ionic cores. This interference effect, together with the different
intensity dependence of the ionization of atomic orbitals, accounts
for the peculiar ionization behavior of O$_2$ comparing to its
companion atom Xe.

This work is supported by NNSF of China (No. 10925420, No.
10904162, No. 11074026 and No. 11104225), the National Basic
Research Program of China Grant No. 2011CB8081002.

\end{document}